\documentclass[sigconf]{acmart}
\usepackage{flushend}
\usepackage{booktabs}
\usepackage{makecell}
\usepackage{colortbl}
\usepackage{subcaption}

\usepackage{booktabs} 
\usepackage{graphicx}
\usepackage{hyperref}
\usepackage{multirow}
\usepackage{hyperref}
\definecolor{darkblue}{rgb}{0.0, 0.0, 0.55}
\definecolor{darkcandyapplered}{rgb}{0.64, 0.0, 0.0}
\hypersetup{
  colorlinks   = true, 
  urlcolor     = darkblue, 
  linkcolor    = darkblue, 
  citecolor   = darkcandyapplered 
}

\usepackage{amsmath,amssymb,amsfonts}
\usepackage{algorithmic}
\usepackage{textcomp}
\usepackage[final]{listings}
\usepackage{float}

\usepackage{graphics} 
\setkeys{Gin}{draft=false}
\usepackage{epsfig} 
\usepackage[colorinlistoftodos,prependcaption,textsize=small]{todonotes}
\presetkeys%
    {todonotes}%
    {inline}{}

\newcommand{\ie}{i.e.,~}
\newcommand{\eg}{e.g.,~}
\newcommand{\etc}{etc.}

\newcommand{\myparagraph}[1]{\vspace{2mm}\noindent\textbf{#1}}

\newcommand{\vspacebfigure}{\vspace{-1.5mm}}   
\newcommand{\vspaceafigure}{\vspace{-2.5mm}}   

\definecolor{javared}{rgb}{0.6,0,0} 
\definecolor{javagreen}{rgb}{0.25,0.5,0.35} 
\definecolor{javapurple}{rgb}{0.5,0,0.35} 
\definecolor{javadocblue}{rgb}{0.25,0.35,0.75} 

\makeatletter
\newenvironment{btHighlight}[1][]
{\begingroup\tikzset{bt@Highlight@par/.style={#1}}\begin{lrbox}{\@tempboxa}}
{\end{lrbox}\bt@HL@box[bt@Highlight@par]{\@tempboxa}\endgroup}

\newcommand\btHL[1][]{%
  \begin{btHighlight}[#1]\bgroup\aftergroup\bt@HL@endenv%
}
\def\bt@HL@endenv{%
  \end{btHighlight}%
  \egroup
}

\newcommand{\bt@HL@box}[2][]{%
  \tikz[#1]{%
    \pgfpathrectangle{\pgfpoint{1pt}{0pt}}{\pgfpoint{\wd #2}{\ht #2}}%
    \pgfusepath{use as bounding box}%
    \node[anchor=base west, fill=orange!30,outer sep=0pt,inner xsep=1pt, inner ysep=0pt, minimum height=\ht\strutbox+1pt,#1]{\raisebox{1pt}{\strut}\strut\usebox{#2}};
  }%
}
\makeatother

\lstdefinestyle{CALCITESQL}{
    language={SQL},basicstyle=\scriptsize\ttfamily,
    moredelim=**[is][\btHL]{`}{`},
    moredelim=**[is][{\btHL[fill=green!30,thin]}]{@}{@},
    keywordstyle=\color{javapurple}\bfseries,
		stringstyle=\color{javared},
		commentstyle=\color{javagreen},
		morecomment=[s][\color{javadocblue}]{/**}{*/},
		numbers=left,
		numberstyle=\tiny\color{black},
		stepnumber=0,
		numbersep=10pt,
		tabsize=4,
		showspaces=false,
		showstringspaces=false
}

\def\BibTeX{{\rm B\kern-.05em{\sc i\kern-.025em b}\kern-.08em
    T\kern-.1667em\lower.7ex\hbox{E}\kern-.125emX}}
\begin{document}

\title[Apache Calcite]{Apache Calcite: A Foundational Framework for Optimized Query Processing Over Heterogeneous Data Sources}

\author{Edmon Begoli}
\affiliation{%
  \institution{Oak Ridge National Laboratory (ORNL)}
  \streetaddress{1 Bethel Valley Rd.}
  \city{Oak Ridge}
  \state{Tennessee}
  \postcode{37831}
  \country{USA}
}
\email{begolie@ornl.gov}

\author{Jes\'us Camacho-Rodr\'iguez}
\affiliation{%
  \institution{Hortonworks Inc.}
  \streetaddress{5470 Great America Pkwy}
  \city{Santa Clara}
  \state{California}
  \postcode{95054}
  \country{USA}
}
\email{jcamacho@hortonworks.com}

\author{Julian Hyde}
\affiliation{%
  \institution{Hortonworks Inc.}
  \streetaddress{5470 Great America Pkwy}
  \city{Santa Clara}
  \state{California}
  \postcode{95054}
  \country{USA}
}
\email{jhyde@hortonworks.com}

\author{Michael J. Mior}
\affiliation{%
  \institution{David R. Cheriton School of Computer Science}
  \institution{University of Waterloo}
  \streetaddress{200 University Avenue West}
  \city{Waterloo}
  \state{Ontario}
  \postcode{N2L 3G1}
  \country{Canada}
}
\email{mmior@uwaterloo.ca}

\author{Daniel Lemire}
\affiliation{%
  \institution{University of Quebec (TELUQ)}
  \streetaddress{5800 Saint-Denis, office 1105}
  \city{Montreal}
  \state{Quebec}
  \postcode{H2S 3L5}
  \country{Canada}
}
\email{lemire@gmail.com}
\renewcommand{\shortauthors}{E. Begoli, J. Camacho-Rodr\'iguez, J. Hyde, M. Mior, and D. Lemire}
\begin{abstract}
Apache Calcite is a foundational software framework that provides query processing, optimization, and query language support to many popular open-source data processing systems such as Apache Hive, Apache Storm, Apache Flink, Druid, and MapD.
Calcite's architecture consists of a modular and extensible query optimizer with hundreds of built-in optimization rules, a query processor capable of processing a variety of query languages, an adapter architecture designed for extensibility, and support for heterogeneous data models and stores (relational, semi-structured, streaming, and geospatial).
This flexible, embeddable, and extensible architecture is what makes Calcite an attractive choice for adoption in big-data frameworks. It is an active project that continues to introduce support for the new types of data sources, query languages, and approaches to query processing and optimization.
\end{abstract}

\copyrightyear{2018} 
\acmYear{2018} 
\setcopyright{licensedusgovmixed}
\acmConference[SIGMOD'18]{2018 International Conference on Management of Data}{June 10--15, 2018}{Houston, TX, USA}
\acmBooktitle{SIGMOD'18: 2018 International Conference on Management of Data, June 10--15, 2018, Houston, TX, USA}
\acmPrice{15.00}
\acmDOI{10.1145/3183713.3190662}
\acmISBN{978-1-4503-4703-7/18/06}

\ccsdesc[500]{Information systems~DBMS engine architectures}

\settopmatter{printacmref=false}

\keywords{Apache Calcite, Relational Semantics, Data Management, Query Algebra, Modular Query Optimization, Storage Adapters}

\maketitle

\section{Introduction}
\label{sec:intro}

Following the seminal System R, conventional relational database engines dominated the data processing landscape.
Yet, as far back as 2005, Stonebraker and \c{C}etintemel~\cite{DBLP:conf/icde/StonebrakerC05} predicted that we would
see the rise a collection of specialized engines such
as column stores, stream processing engines, text search engines, and so forth.
They argued that specialized engines can offer more cost-effective performance and that they would bring the end of the ``one size fits all'' paradigm. Their vision seems today more relevant than ever.
Indeed, many specialized open-source data systems have since become popular such as Storm~\cite{website:Storm} and Flink~\cite{website:Flink} (stream processing), Elasticsearch~\cite{website:Elastic} (text search), Apache Spark~\cite{website:Spark}, Druid~\cite{website:Druid}, \etc

As organizations have invested in data processing systems tailored towards their specific needs, two overarching problems have arisen:

\begin{itemize}
\item The developers of such specialized systems have encountered related problems, such as query optimization~\cite{DBLP:conf/sigmod/ArmbrustXLHLBMK15,DBLP:conf/sigmod/HuaiCGHHOPYL014}
or the need to support query languages such as SQL and related extensions (\eg streaming queries~\cite{DBLP:journals/cacm/Hyde10}) as well as language-integrated queries inspired by LINQ~\cite{DBLP:conf/sigmod/MeijerBB06}.
Without a unifying framework, having multiple engineers independently develop similar optimization logic and language support wastes engineering effort.
\item Programmers using these specialized systems often have to integrate several of them together.
An organization might rely on Elasticsearch, Apache Spark, and Druid.
We need to build systems capable of supporting optimized queries across heterogeneous data sources~\cite{8091081}.
\end{itemize}

Apache Calcite was developed to solve these problems.
It is a complete query processing system that provides much of the common functionality---query execution, optimization, and query languages---required by any database management system, except for data storage and management, which are left to specialized engines.
Calcite was quickly adopted by Hive, Drill~\cite{website:Drill}, Storm, and many other data processing engines, providing them with advanced query
optimizations and query languages.\footnote{\url{http://calcite.apache.org/docs/powered_by}}
For example, Hive~\cite{website:Hive} is a popular data warehouse project built on top of Apache Hadoop. As Hive moved from its batch processing roots towards an interactive SQL query answering platform, it became clear that the project needed a powerful optimizer at its core.
Thus, Hive adopted Calcite as its optimizer and their integration has been growing since.
Many other projects and products have followed suit, including Flink, MapD~\cite{website:MapDblog}, \etc

Furthermore, Calcite enables cross-platform optimization by exposing a common interface to multiple systems.
To be efficient, the optimizer needs to reason globally, \eg make decisions across different systems about materialized view selection.

Building a common framework does not come without challenges. In particular, the framework needs to be extensible and flexible enough to accommodate the different types of systems requiring integration.

We believe that the following features have contributed to  Calcite's wide adoption in the open source community and industry:

\begin{itemize}
	\item\textbf{Open source friendliness.} Many of the major data processing platforms of the last decade have been either open source or largely based on open source.
Calcite is an open-source framework, backed by the Apache Software Foundation (ASF)~\cite{asf:website}, which provides  the means to collaboratively develop the project.
Furthermore, the software is written in Java making it easier to interoperate with many of the latest data processing systems~\cite{website:Drill,website:Flink,website:Hive,website:Kylin,website:Samza,website:MapDblog} that are often written themselves in Java (or in the JVM-based Scala), especially those in the Hadoop ecosystem.
	\item\textbf{Multiple data models.} Calcite provides support for query optimization and query languages using both streaming and conventional data processing paradigms.
Calcite treats streams as time-ordered sets of records or events that
are not persisted to the disk as they would be in conventional data processing systems.
	\item\textbf{Flexible query optimizer.} Each component of the optimizer is pluggable and extensible, ranging from rules to cost models.
In addition, Calcite includes support for multiple planning engines. Hence, the optimization  can be broken down into phases  handled by different optimization engines depending on which one is best suited for the stage.
	\item\textbf{Cross-system support.} The Calcite framework can run and optimize queries across multiple query processing systems and database backends.
	\item\textbf{Reliability.} Calcite is reliable, as its wide adoption over many years has led to exhaustive testing of the platform.
Calcite also contains an extensive test suite validating all components of the system including query optimizer rules and integration with backend data sources.
	\item\textbf{Support for SQL and its extensions.} Many systems do not provide their own query language, but rather prefer to rely on existing ones such as SQL\@. For those, Calcite provides support for ANSI standard SQL, as well as various SQL dialects and extensions, \eg for expressing queries on streaming or nested data. In addition, Calcite includes a driver conforming to the standard Java API (JDBC).
\end{itemize}

The remainder  is organized as follows. Section~\ref{sec:related} discusses related work. Section~\ref{sec:archi} introduces Calcite's architecture and its main components. Section~\ref{sec:relexprs} describes the relational algebra at the core of Calcite.
Section~\ref{sec:adapters} presents Calcite's adapters, an abstraction to define how to read external data sources.
In turn, Section~\ref{sec:optimizer} describes Calcite's optimizer and its main features, while Section~\ref{sec:calcite_extensions} presents the extensions to handle different query processing paradigms.
Section~\ref{sec:action} provides an overview of the data processing systems already using Calcite.
Section~\ref{sec:future} discusses possible future extensions for the framework before we conclude in Section~\ref{sec:conclusion}.


\section{Related Work}
\label{sec:related}
Though  Calcite is currently the most widely adopted optimizer for big-data analytics in the Hadoop ecosystem, many of the ideas that lie behind it are not novel.
For instance, the query optimizer builds on ideas from the Volcano~\cite{Graefe93thevolcano} and Cascades~\cite{DBLP:journals/debu/Graefe95a} frameworks, incorporating other widely used optimization techniques such as materialized view rewriting~\cite{DBLP:conf/icde/ChaudhuriKPS95, DBLP:conf/sigmod/GoldsteinL01,DataCubes}.
There are other systems that try to fill a similar role to Calcite.


Orca~\cite{Soliman:2014:OMQ:2588555.2595637} is a modular query optimizer used in data management products such as Greenplum and HAWQ\@.
Orca decouples the optimizer from the query execution engine by implementing a framework for exchanging information between the two known as \emph{Data eXchange Language}.
Orca also provides tools for verifying the correctness and performance of generated query plans.
In contrast to Orca, Calcite can be used as a standalone query execution engine that federates multiple storage and processing backends, including pluggable planners, and optimizers.

Spark SQL~\cite{Armbrust2015} extends Apache Spark to support SQL query execution which can also execute queries over multiple data sources as in Calcite.
However, although the Catalyst optimizer in Spark SQL also attempts to minimize query execution cost, it lacks the dynamic programming approach used by Calcite and risks falling into local minima.

Algebricks~\cite{Borkar:2015:ADM:2806777.2806941} is a query compiler architecture that provides a data model agnostic algebraic layer and compiler framework for big data query processing.
High-level languages are compiled to Algebricks logical algebra.
Algebricks then generates an optimized job targeting the Hyracks parallel processing backend.
While Calcite shares a modular approach with Algebricks, Calcite also includes a support for cost-based optimizations.
In the current version of Calcite, the query optimizer architecture uses dynamic programming-based planning based on Volcano~\cite{Graefe93thevolcano} with extensions for multi-stage optimizations as in Orca~\cite{Soliman:2014:OMQ:2588555.2595637}. Though in principle Algebricks could support multiple processing backends (e.g., Apache Tez, Spark), Calcite has provided well-tested support for diverse backends for many years.

Garlic~\cite{Carey1995} is a heterogeneous data management system which represents data from multiple systems under a unified object model.
However, Garlic does not support query optimization across different systems and relies on each system to optimize its own queries.

FORWARD~\cite{fu2011sql} is a federated query processor that implements a superset of SQL called SQL++~\cite{ong2014sql++}.
SQL++ has a semi-structured data model that integrate both JSON and relational data models whereas Calcite supports semi-structured data models by representing them in the relational data model during query planning.
FORWARD decomposes federated queries written in SQL++ into subqueries and executes them on the underlying databases according to the query plan.
The merging of data happens inside the FORWARD engine.

Another federated data storage and processing system is BigDAWG, which abstracts a wide spectrum of data models including relational, time-series and streaming.
The unit of abstraction in BigDAWG is called an \emph{island of information}.
Each island of information has a query language, data model and connects to one or more storage systems.
Cross storage system querying is supported within the boundaries of a single island of information.
Calcite instead provides a unifying relational abstraction which allows querying across backends with different data models.

Myria is a general-purpose engine for big data analytics, with advanced support for the Python language~\cite{DBLP:conf/sigmod/HalperinACCKMORWWXBHS14}. It
produces  query plans for other backend engines such as Spark and PostgreSQL\@.

\section{Architecture}
\label{sec:archi}

Calcite contains many of the pieces that comprise a typical database management system. However, it omits some key components, \eg storage of data, algorithms to process data, and a repository for storing metadata. These omissions are deliberate: it makes Calcite an excellent choice for mediating between applications having one or more data storage locations and using multiple data processing engines. It is also a solid foundation for building bespoke data processing systems.

\begin{figure}[t]
\centering
\includegraphics[width=0.94\columnwidth]{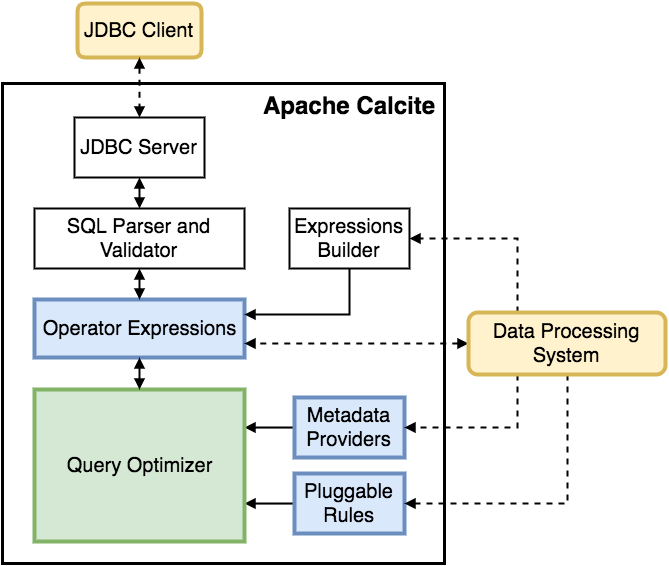}
\vspacebfigure\caption{Apache Calcite architecture and interaction.\label{fig:arch}}\vspaceafigure
\end{figure}

Figure~\ref{fig:arch} outlines the main components of Calcite's architecture.
Calcite's optimizer uses a tree of relational operators as its internal representation.
The optimization engine primarily consists of three components: rules, metadata providers, and planner engines.
We discuss these components in more detail in Section~\ref{sec:optimizer}.
In the figure, the dashed lines represent possible external interactions with the framework.
There are different ways to interact with Calcite.

First, Calcite contains a query parser and validator that can translate a SQL query to a tree of relational operators. As Calcite does not contain a \textit{storage layer}, it provides a mechanism to define table schemas and views in external storage engines via \textit{adapters} (described in Section~\ref{sec:adapters}), so it can be used on top of these engines.

Second, although Calcite provides optimized SQL support to systems that need such database language support, it also provides optimization support
to systems that already have their own language parsing and interpretation:
\begin{itemize}
\item Some systems support SQL queries, but without or with limited query optimization.
For example, both Hive and Spark initially offered support for the SQL language, but they did not include an optimizer.
For such cases, once the query has been optimized, Calcite can translate the relational expression back to SQL\@.
This feature allows Calcite to work as a stand-alone system on top of any data management system with a SQL interface, but no optimizer.
\item The Calcite architecture is not only tailored towards optimizing SQL queries.
It is common that data processing systems choose to use their own parser for their own query language. Calcite can help optimize these queries as well.
Indeed, Calcite also allows operator trees to be easily constructed by directly instantiating relational operators.
One can use the built-in \textit{relational expressions builder} interface.
For instance, assume that we want to express the following Apache Pig~\cite{website:Pig} script using the expression builder:

\begin{lstlisting}
emp = LOAD 'employee_data' AS (deptno, sal);
emp_by_dept = GROUP emp by (deptno);
emp_agg = FOREACH emp_by_dept GENERATE GROUP as deptno,
    COUNT(emp.sal) AS c, SUM(emp.sal) as s;
dump emp_agg;
\end{lstlisting}

The equivalent expression looks as follows:

\begin{lstlisting}[language=Java]
final RelNode node = builder
  .scan("employee_data")
  .aggregate(builder.groupKey("deptno"),
             builder.count(false, "c"),
             builder.sum(false, "s", builder.field("sal")))
  .build();
\end{lstlisting}

This interface exposes the main constructs necessary for building relational expressions.
After the optimization phase is finished, the application can retrieve the optimized relational expression which can then be mapped back to the system's query processing unit.
\end{itemize}

\section{Query Algebra}
\label{sec:relexprs}

\myparagraph{Operators.} Relational algebra~\cite{DBLP:journals/cacm/Codd70} lies at the core of Calcite.
In addition to the operators that express the most common data manipulation operations, such as \textit{filter}, \textit{project}, \textit{join} \etc, Calcite includes additional operators that meet different purposes, \eg being able to concisely represent complex operations, or recognize optimization opportunities more efficiently.

For instance, it has become common for OLAP, decision making, and streaming applications to use window definitions to express complex analytic functions such as moving average of a quantity over a time period or number or rows.  Thus, Calcite introduces a \textit{window} operator that encapsulates the window definition, \ie upper and lower bound, partitioning \etc, and the aggregate functions to execute on each window.

\myparagraph{Traits.} Calcite does not use different entities to represent logical and physical operators.
Instead, it describes the physical properties associated with an operator using \textit{traits}.
These traits help the optimizer evaluate the cost of different alternative plans.
Changing a trait value does not change the logical expression being evaluated, \ie the rows produced by the given operator will still be the same.

During optimization, Calcite tries to enforce certain traits on relational expressions, \eg the sort order of certain columns.
Relational operators can implement a \textit{converter} interface that indicates how to convert traits of an expression from one value to another.

Calcite includes common traits that describe the physical properties of the data produced by a relational expression, such as \textit{ordering}, \textit{grouping}, and \textit{partitioning}.
Similar to the {SCOPE} optimizer~\cite{DBLP:conf/icde/ZhouLC10}, the Calcite optimizer can reason about these properties and exploit them to find plans that avoid unnecessary operations.
For example, if the input to the sort operator is already correctly ordered---possibly because this is the same order used for rows in the backend system---then the sort operation can be removed.

In addition to these properties, one of the main features of Calcite is the \emph{calling convention} trait.
Essentially, the trait represents the data processing system where the expression will  be executed.
Including the calling convention as a trait allows Calcite to meet its goal of optimizing transparently queries whose execution might span over different engines \ie the convention will be treated as any other physical property.

\begin{figure}[t]
\centering
\includegraphics[width=1\columnwidth]{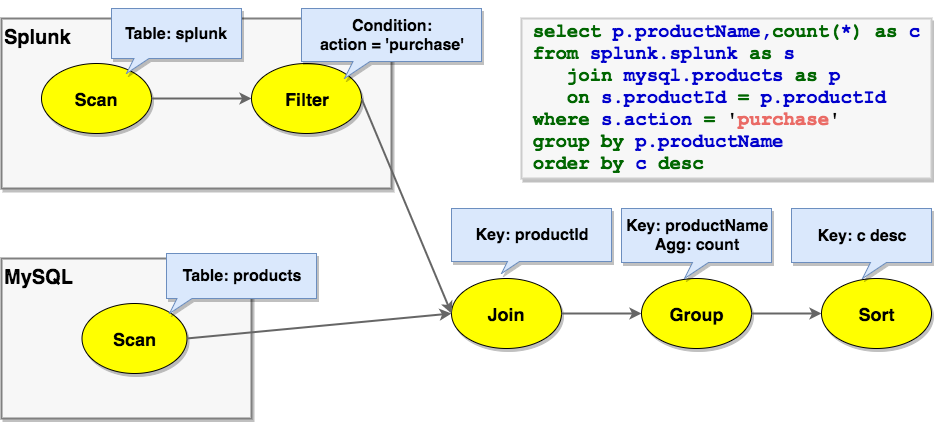}
\vspacebfigure\caption{A Query Optimization Process.\label{fig:optimizer}}\vspaceafigure
\end{figure}

For example, consider joining a \textit{Products} table held in MySQL to an \textit{Orders} table held in Splunk (see Figure~\ref{fig:optimizer}).
Initially, the scan of \textit{Orders} takes place in the \textit{splunk} convention and the scan of \textit{Products} is in the \textit{jdbc-mysql} convention. The tables have to be scanned inside their respective engines.
The join is in the \textit{logical} convention, meaning that no implementation has been chosen.
Moreover, the SQL query in Figure~\ref{fig:optimizer} contains a filter (\texttt{where} clause) which is pushed into \textit{splunk} by an adapter-specific rule (see Section~\ref{sec:adapters}).
One possible implementation is to use Apache Spark as an external engine: the join is converted to \textit{spark} convention, and its inputs are converters from \textit{jdbc-mysql} and \textit{splunk} to \textit{spark} convention.
But there is a more efficient implementation: exploiting the fact that Splunk can perform lookups into MySQL via ODBC, a planner rule pushes the join through the \textit{splunk}-to-\textit{spark} converter, and the join is now in \textit{splunk} convention, running inside the Splunk engine.


\section{Adapters}
\label{sec:adapters}
\begin{figure}[t]
\centering
\includegraphics[width=0.95\columnwidth,trim={5 5 5 5},clip]{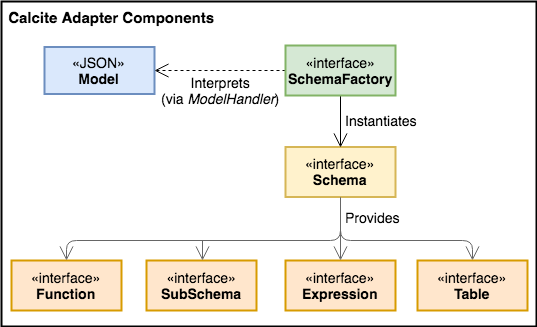}
\vspacebfigure\caption{Calcite's Data Source Adapter Design.\label{fig:adapter}}\vspaceafigure
\end{figure}

An adapter is an architectural pattern that defines how Calcite incorporates diverse data sources for general access.
Figure~\ref{fig:adapter} depicts its components. Essentially, an adapter consists of a model, a schema, and a schema factory.
The \textit{model} is a specification of the physical properties of the data source being accessed.
A \textit{schema} is the definition of the data (format and layouts) found in the model.
The data itself is physically accessed via tables.
Calcite interfaces with the tables defined in the adapter to read the data as the query is being executed. The adapter may define a set of rules that are added to the planner. For instance, it typically includes rules to convert various types of logical relational expressions to the corresponding relational expressions of the adapter's convention.
The \textit{schema factory} component acquires the metadata information from the model and generates a schema.


As discussed in Section~\ref{sec:relexprs}, Calcite uses a physical trait known as the \textit{calling convention} to identify relational operators which correspond to a specific database backend.
These physical operators implement the access paths for the underlying tables in each adapter.
When a query is parsed and converted to a relational algebra expression, an operator is created for each table representing a scan of the data on that table.
It is the minimal interface that an adapter must implement.
If an adapter implements the table scan operator, the Calcite optimizer is then able to use client-side operators such as sorting, filtering, and joins to execute arbitrary SQL queries against these tables.

This table scan operator contains the necessary information the adapter requires to issue the scan to the adapter's backend database.
To extend the functionality provided by adapters, Calcite defines an \emph{enumerable} calling convention.
Relational operators with the enumerable calling convention simply operate over tuples  via an iterator interface.
This calling convention allows Calcite to implement operators which may not be available in each adapter's backend.
For example, the \texttt{EnumerableJoin} operator implements joins by collecting rows from its child nodes and joining on the desired attributes.

For queries which only touch a small subset of the data in a table, it is inefficient for Calcite to  enumerate all tuples.
Fortunately, the same rule-based optimizer can be used to implement adapter-specific rules for optimization.
For example, suppose a query involves filtering and sorting on a table.
An adapter which can perform filtering on the backend can implement a rule which matches a \texttt{LogicalFilter} and converts it to the adapter's calling convention.
This rule converts the \texttt{LogicalFilter} into another \texttt{Filter} instance.
This new \texttt{Filter} node has a lower associated cost that allows Calcite to optimize queries across adapters.

The use of adapters is a powerful abstraction that enables not only optimization of queries for a specific backend, but also across multiple backends.
Calcite is able to answer queries involving tables across multiple backends by pushing down all possible logic to each backend and then performing joins and aggregations on the resulting data.
Implementing an adapter can be as simple as providing a table scan operator or it can involve the design of many advanced optimizations.
Any expression represented in the relational algebra can be pushed down to adapters with optimizer rules.


\section{Query Processing and Optimization}
\label{sec:optimizer}
The query optimizer is the main component in the framework. Calcite optimizes queries by repeatedly applying planner rules to a relational expression. A cost model guides the process, and the planner engine tries to generate an alternative expression that has the same semantics as the original but a lower cost.

Every component in the optimizer is extensible. Users can add relational operators, rules, cost models, and statistics.

\myparagraph{Planner rules.}
Calcite includes a set of planner rules to transform expression trees.
In particular, a rule matches a given pattern in the tree and executes a transformation that preserves semantics of that expression.
Calcite includes several hundred optimization rules.
However, it is rather common for data processing systems relying on Calcite for optimization to include their own rules to allow specific rewritings.

For example, Calcite provides an adapter for Apache Cassandra~\cite{Cassandra}, a wide column store which partitions data by a subset of columns in a table and then within each partition, sorts rows based on another subset of columns. As discussed in Section~\ref{sec:adapters}, it is beneficial for adapters to push down as much query processing as possible to each backend for efficiency.
A rule to push a \texttt{Sort} into Cassandra must check two conditions:

\begin{enumerate}
\item the table has been previously filtered to a single partition (since rows are only sorted within a partition) and
\item the sorting of partitions in Cassandra has some common prefix with the required sort.
\end{enumerate}

This requires that a \texttt{LogicalFilter} has been rewritten to a \texttt{CassandraFilter} to ensure the partition filter is pushed down to the database.
The effect of the rule is simple (convert a \texttt{LogicalSort} into a \texttt{CassandraSort}) but the flexibility in rule matching enables backends to push down operators even in complex scenarios.

\begin{figure}
\centering
\subcaptionbox{Before\label{fig:filter-into-join-before}}{%
\includegraphics[width=0.45\columnwidth]{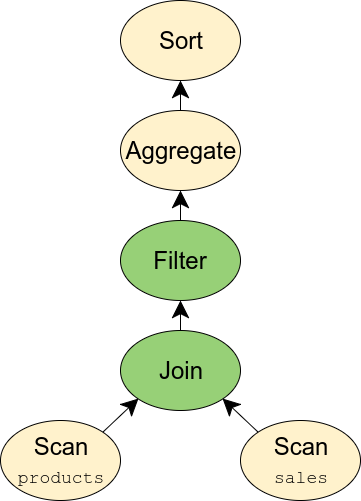}
}
\subcaptionbox{After\label{fig:filter-into-join-after}}{%
\includegraphics[width=0.45\columnwidth]{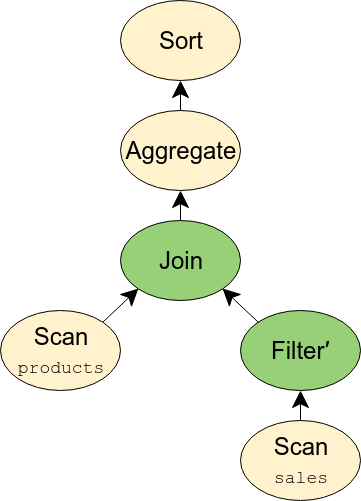}
}
\vspacebfigure\caption{\texttt{FilterIntoJoinRule} application.\label{fig:filter-into-join}}\vspaceafigure
\end{figure}

For an example of a rule with more complex effects, consider the following query:

\begin{lstlisting}[style=CALCITESQL]
SELECT products.name, COUNT(*)
FROM sales JOIN products USING (productId)
WHERE sales.discount IS NOT NULL
GROUP BY products.name
ORDER BY COUNT(*) DESC;
\end{lstlisting}

The query corresponds to the relational algebra expression presented in Figure~\ref{fig:filter-into-join-before}.
Because the \texttt{WHERE} clause only applies to the \texttt{sales} table,  we can move the filter before the join as in Figure~\ref{fig:filter-into-join-after}.
This optimization can significantly reduce query execution time since we do not need to perform the join for rows which do match the predicate.
Furthermore, if the \texttt{sales} and \texttt{products} tables were contained in separate backends, moving the filter before the join also potentially enables an adapter to push the filter into the backend.
Calcite implements this optimization via \texttt{FilterIntoJoinRule} which matches a filter node with a join node as a parent and checks if the filter can be performed by the join.
This optimization illustrates the flexibility of the Calcite approach to optimization.

\myparagraph{Metadata providers.} Metadata is an important part of Calcite's optimizer, and it serves two main purposes: ($i$)~guiding the planner towards the goal of reducing the cost of the overall query plan, and ($ii$)~providing information to the rules while they are being applied.

Metadata providers are responsible for supplying that information to the optimizer. In particular, the default metadata providers implementation in Calcite contains functions that return the overall cost of executing a subexpression in the operator tree, the number of rows and the data size of the results of that expression, and the maximum degree of parallelism with which it can be executed.
In turn, it can also provide information about the plan structure, \eg filter conditions that are present below a certain tree node.

Calcite provides interfaces that allow data processing systems to plug their metadata information into the framework. These systems may choose to write providers that override the existing functions, or provide their own new metadata functions that might be used during the optimization phase. However, for many of them, it is sufficient to provide statistics about their input data, \eg number of rows and size of a table, whether values for a given column are unique \etc, and Calcite will do the rest of the work by using its default implementation.

As the metadata providers are pluggable, they are compiled and instantiated at runtime using Janino~\cite{website:Janino}, a Java lightweight compiler. Their implementation includes a cache for metadata results, which yields significant performance improvements, \eg when we need to compute multiple types of metadata such as \textit{cardinality}, \textit{average row size}, and \textit{selectivity} for a given join, and all these computations rely on the cardinality of their inputs.


\myparagraph{Planner engines.} The main goal of a planner engine is to trigger the rules provided to the engine until it reaches a given objective.
At the moment, Calcite provides two different engines. 
New engines are pluggable in the framework. 

The first one, \textit{a cost-based planner engine}, triggers the input rules with the goal of reducing the overall expression cost.
The engine uses a dynamic programming algorithm, similar to Volcano~\cite{Graefe93thevolcano}, to create and track different alternative plans created by firing the rules given to the engine.
Initially, each expression is registered with the planner, together with a digest based on the expression attributes and its inputs. When a rule is fired on an expression $e_1$ and the rule produces a new expression $e_2$, the planner will add $e_2$ to the set of equivalence expressions $S_a$ that $e_1$ belongs to. In addition, the planner generates a digest for the new expression, which is compared with those previously registered in the planner. If a similar digest associated with an expression $e_3$ that belongs to a set $S_b$ is found, the planner has found a duplicate and hence will merge $S_a$ and $S_b$ into a new set of equivalences.
The process continues until the planner reaches a configurable fix point. In particular, it can ($i$)~exhaustively explore the search space until all rules have been applied on all expressions, or ($ii$)~use a heuristic-based approach to stop the search when the plan cost has not improved by more than a given threshold $\delta$ in the last planner iterations.
The cost function that allows the optimizer to decide which plan to choose is supplied through metadata providers. The default cost function implementation combines estimations for CPU, IO, and memory resources used by a given expression.

The second engine is an \textit{exhaustive planner}, which triggers rules exhaustively until it generates an expression that is no longer modified by any rules. This planner is useful to quickly execute rules without taking into account the cost of each expression.

Users may choose to use one of the existing planner engines depending on their concrete needs, and switching from one to another, when their system requirements change, is straightforward. Alternatively, users may choose to generate \textit{multi-stage optimization} logic, in which different sets of rules are applied in consecutive phases of the optimization process.
Importantly, the existence of two planners allows Calcite users to reduce the overall optimization time by guiding the search for different query plans.

\myparagraph{Materialized views.} One of the most powerful techniques to accelerate query processing in data warehouses is the precomputation of relevant summaries or \textit{materialized views}.
Multiple Calcite adapters and projects relying on Calcite have their own notion of materialized views.
For instance, Cassandra allows the user to define materialized views based on existing tables which are automatically maintained by the system.

These engines expose their materialized views to Calcite.
The optimizer then has the opportunity to rewrite incoming queries to use these views instead of the original tables.
In particular, Calcite provides an implementation of two different materialized view-based rewriting algorithms.

The first approach is based on \textit{view substitution}~\cite{DBLP:conf/icde/ChaudhuriKPS95, DBLP:conf/sigmod/GoldsteinL01}.
The aim is to substitute part of the relational algebra tree with an equivalent expression which makes use of a materialized view, and the algorithm proceeds as follows: ($i$)~the scan operator over the materialized view and the materialized view definition plan are registered with the planner, and ($ii$)~transformation rules that try to unify expressions in the plan are triggered.
Views do not need to exactly match expressions in the query being replaced, as the rewriting algorithm in Calcite can produce partial rewritings that include additional operators to compute the desired expression, \eg filters with residual predicate conditions.

The second approach is based on \textit{lattices}~\cite{DataCubes}. Once the data sources are declared to form a lattice, Calcite represents each of the materializations as a \textit{tile} which in turn can be used by the optimizer to answer incoming queries. On the one hand, the rewriting algorithm is especially efficient in matching expressions over data sources organized in a star schema, which are common in OLAP applications. On the other hand, it is more restrictive than view substitution, as it imposes restrictions on the underlying schema.

\section{Extending Calcite}
\label{sec:calcite_extensions}

As we have mentioned in the previous sections, Calcite is not only tailored towards SQL processing.
In fact, Calcite provides extensions to SQL expressing queries over other data abstractions, such as semi-structured, streaming and geospatial data.
Its internal operators adapt to these queries. 
In addition to extensions to SQL, Calcite also includes a language-integrated query language.
We describe these extensions throughout this section and provide some examples.

\subsection{Semi-structured Data}

Calcite supports several complex column data types that enable a hybrid of relational and semi-structured data to be stored in tables.
Specifically, columns can be of type \texttt{ARRAY}, \texttt{MAP}, or \texttt{MULTISET}.
Furthermore, these complex types can be nested so it is possible for example, to have a \texttt{MAP} where the values are of type \texttt{ARRAY}.
Data within the \texttt{ARRAY} and \texttt{MAP} columns (and nested data therein) can be extracted using the \texttt{[]} operator.
The specific type of values stored in any of these complex types need not be predefined.

For example, Calcite contains an adapter for MongoDB~\cite{website:Mongo}, a document store which stores documents consisting of data roughly equivalent to JSON documents.
To expose MongoDB data to Calcite, a table is created for each document collection with a single column named \texttt{\_MAP}: a map from document identifiers to their data.
In many cases, documents can be expected to have a common structure.
A collection of documents representing zip codes may each contain columns with a city name, latitude and longitude.
It can be useful to expose this data as a relational table.
In Calcite, this is achieved by creating a view after extracting the desired values and casting them to the appropriate type:

\begin{lstlisting}[style=CALCITESQL]
SELECT CAST(_MAP['city'] AS varchar(20)) AS city,
  CAST(_MAP['loc'][0] AS float) AS longitude,
  CAST(_MAP['loc'][1] AS float) AS latitude
FROM mongo_raw.zips;
\end{lstlisting}
With views over semi-structured data defined in this manner, it becomes easier to manipulate data from different semi-structured sources in tandem with relational data.


\subsection{Streaming}
Calcite provides first-class support for streaming queries~\cite{DBLP:journals/cacm/Hyde10} based on a set of streaming-specific extensions to standard SQL\@, namely \textit{STREAM} extensions, windowing extensions, implicit references to streams via window expressions in joins, and others.
These extensions were inspired by the Continuous Query Language~\cite{ilprints758} while also trying to integrate effectively with standard SQL\@.
The primary extension, the \texttt{STREAM} directive tells the system that the user is interested in incoming records, not existing ones.

\begin{lstlisting}[style=CALCITESQL]
SELECT @STREAM@ rowtime, productId, units
FROM Orders
WHERE units > 25;
\end{lstlisting}

In the absence of the \texttt{STREAM} keyword when querying a stream, the query becomes a regular relational query, indicating the system should process existing records which have already been received from a stream, not the incoming ones.

Due to the inherently unbounded nature of streams, windowing is used to unblock blocking operators such as aggregates and joins.
Calcite's streaming extensions use SQL analytic functions to express sliding and cascading window aggregations, as shown in the following example.

\begin{lstlisting}[style=CALCITESQL]
SELECT STREAM rowtime,
  productId,
  units,
  SUM(units) @OVER (ORDER BY rowtime@
  	@PARTITION BY productId@
  	@RANGE INTERVAL '1' HOUR PRECEDING)@ unitsLastHour
FROM Orders;
\end{lstlisting}
Tumbling, hopping and session windows\footnote{Tumbling, hopping, sliding, and session windows are different schemes for grouping of the streaming events~\cite{mendes2009performance}.} are enabled by the \texttt{TUMBLE}, \texttt{HOPPING}, \texttt{SESSION} functions and related utility functions such as \texttt{TUMBLE\_END} and \texttt{HOP\_END} that can be used respectively in \texttt{GROUP BY} clauses and projections.

\begin{lstlisting}[style=CALCITESQL]
SELECT STREAM
  @TUMBLE_END(rowtime, INTERVAL '1' HOUR)@ AS rowtime,
  productId,
  COUNT(*) AS c,
  SUM(units) AS units
FROM Orders
GROUP BY @TUMBLE(rowtime, INTERVAL '1' HOUR)@, productId;
\end{lstlisting}

Streaming queries involving window aggregates require the presence of monotonic or quasi-monotonic expressions in the \texttt{GROUP BY} clause or in the \texttt{ORDER BY} clause in case of sliding and cascading window queries.

Streaming queries which involve more complex stream-to-stream joins can be expressed using an implicit (time) window expression in the \texttt{JOIN} clause.

\begin{lstlisting}[style=CALCITESQL]
SELECT STREAM o.rowtime, o.productId, o.orderId,
  s.rowtime AS shipTime
FROM Orders AS o
JOIN Shipments AS s
  ON o.orderId = s.orderId
  AND @s.rowtime BETWEEN o.rowtime AND@
  @o.rowtime + INTERVAL '1' HOUR@;
\end{lstlisting}

In the case of an implicit window, Calcite's query planner validates that the  expression is monotonic.

\subsection{Geospatial Queries}

Geospatial support is preliminary in Calcite, but is being implemented using Calcite's relational algebra.
The core of this implementation consists in adding a new \texttt{GEOMETRY} data type which encapsulates different geometric objects such as points, curves, and polygons.
It is expected  that Calcite will be fully compliant with the OpenGIS Simple Feature Access~\cite{OpenGIS:SFA} specification which defines a standard for SQL interfaces to access geospatial data.  An example query finds the country which contains the city of Amsterdam:
\begin{lstlisting}[style=CALCITESQL]
SELECT name FROM (
  SELECT name,
    ST_GeomFromText('POLYGON((4.82 52.43, 4.97 52.43, 4.97 52.33,
      4.82 52.33, 4.82 52.43))') AS "Amsterdam",
    ST_GeomFromText(boundary) AS "Country"
  FROM country
) WHERE ST_Contains("Country", "Amsterdam");
\end{lstlisting}

%



\subsection{Language-Integrated Query for Java}

Calcite can be used to query multiple data sources, beyond just relational databases. But it also aims to support more than just the SQL language.
Though SQL remains the primary database language, many programmers favour
language-integrated languages like LINQ~\cite{DBLP:conf/sigmod/MeijerBB06}. Unlike SQL embedded within Java or C++ code, language-integrated query languages allow the programmer to write all of her code using a single language.
Calcite provides  Language-Integrated Query for Java (or LINQ4J, in short) which closely follows the convention set forth by Microsoft's LINQ for the .NET languages.

\section{Industry and academia adoption}
\label{sec:action}

Calcite enjoys wide adoption, specially among open-source projects used in industry. As Calcite provides certain integration flexibility, these projects have chosen to either ($i$)~embed Calcite within their core, \ie use it as a library, or ($ii$)~implement an adapter to allow Calcite to federate query processing. In addition, we see a growing interest in the research community to use Calcite as the cornerstone of the development of data management projects. In the following, we describe how different systems are using Calcite.

{%
\renewcommand{\arraystretch}{1.5}
\begin{table*}[th]
\centering
{\begin{tabular}{cccccc} \toprule
\textbf{System} & \textbf{Query Language} & \makecell{\textbf{JDBC}\\ \textbf{Driver}} & \makecell{\textbf{SQL Parser}\\ \textbf{and Validator}} & \makecell{\textbf{Relational}\\ \textbf{Algebra}} & \textbf{Execution Engine} \\
\midrule
Apache Drill & SQL + extensions       & \cellcolor{green!10}\checkmark & \cellcolor{green!10}\checkmark & \cellcolor{green!10}\checkmark & Native \\
Apache Hive & SQL + extensions   &   &   & \cellcolor{green!10}\checkmark   & Apache Tez, Apache Spark \\
Apache Solr & SQL & \cellcolor{green!10}\checkmark & \cellcolor{green!10}\checkmark & \cellcolor{green!10}\checkmark & Native, Enumerable, Apache Lucene \\
Apache Phoenix & SQL & \cellcolor{green!10}\checkmark & \cellcolor{green!10}\checkmark & \cellcolor{green!10}\checkmark & Apache HBase  \\
Apache Kylin & SQL        & \cellcolor{green!10}\checkmark & \cellcolor{green!10}\checkmark &   & Enumerable, Apache HBase  \\
Apache Apex & Streaming SQL & \cellcolor{green!10}\checkmark & \cellcolor{green!10}\checkmark & \cellcolor{green!10}\checkmark & Native \\
Apache Flink & Streaming SQL & \cellcolor{green!10}\checkmark & \cellcolor{green!10}\checkmark & \cellcolor{green!10}\checkmark & Native \\
Apache Samza & Streaming SQL & \cellcolor{green!10}\checkmark & \cellcolor{green!10}\checkmark & \cellcolor{green!10}\checkmark & Native  \\
Apache Storm & Streaming SQL & \cellcolor{green!10}\checkmark & \cellcolor{green!10}\checkmark & \cellcolor{green!10}\checkmark & Native  \\
MapD~\cite{website:MapD} & SQL & & \cellcolor{green!10}\checkmark & \cellcolor{green!10}\checkmark & Native  \\
Lingual~\cite{website:Lingual} & SQL      &   & \cellcolor{green!10}\checkmark & \cellcolor{green!10}\checkmark & Cascading  \\
Qubole Quark~\cite{website:Qubole} & SQL & \cellcolor{green!10}\checkmark & \cellcolor{green!10}\checkmark & \cellcolor{green!10}\checkmark & Apache Hive, Presto   \\
\bottomrule
\end{tabular}}
\caption{List of systems that embed Calcite.\label{tab:systems}}\vspaceafigure
\end{table*}
}

\subsection{Embedded Calcite}

Table~\ref{tab:systems} provides a list of software that incorporates Calcite as a library, including ($i$)~the query language interface that they expose to users, ($ii$)~whether they use Calcite's JDBC driver (called \textit{Avatica}), ($iii$)~whether they use the SQL parser and validator included in Calcite, ($iv$)~whether they use Calcite's query algebra to represent their operations over data, and ($v$)~the engine that they rely on for execution, \eg their own native engine, Calcite's operators (referred to as \textit{enumerable}), or any other project. 

Drill~\cite{website:Drill} is a flexible data processing engine based on the Dremel system~\cite{DBLP:journals/pvldb/MelnikGLRSTV10} that internally uses a schema-free JSON document data model.
Drill uses its own dialect of SQL that includes extensions to express queries on semi-structured data, similar to SQL++~\cite{ong2014sql++}.

Hive~\cite{website:Hive} first became popular as a SQL interface on top of the MapReduce programming model~\cite{hive}. It has since moved towards being an interactive SQL query answering engine, adopting Calcite as its rule and cost-based optimizer.
Instead of relying on Calcite's JDBC driver, SQL parser and validator, Hive uses its own implementation of these components.
The query is then translated into Calcite operators, which after optimization are translated into Hive's physical algebra.
Hive operators can be executed by multiple engines, the most popular being Apache Tez~\cite{DBLP:conf/sigmod/SahaSSVMC15,website:Tez} and Apache Spark~\cite{Zaharia:sparksystem,website:Spark}.

Apache Solr~\cite{website:Solr} is a popular full-text distributed search platform built on top of the Apache Lucene library~\cite{website:Lucene}. Solr exposes multiple query interfaces to users, including REST-like HTTP/XML and JSON APIs. In addition, Solr integrates with Calcite to provide SQL compatibility.

Apache Phoenix~\cite{website:Phoenix} and Apache Kylin~\cite{website:Kylin} both work on top of Apache HBase~\cite{website:HBase}, a distributed key-value store modeled after Bigtable~\cite{DBLP:conf/osdi/ChangDGHWBCFG06}.
In particular, Phoenix provides a SQL interface and orchestration layer to query HBase.
Kylin focuses on OLAP-style SQL queries instead, building cubes that are declared as materialized views and stored in HBase, and hence allowing Calcite's optimizer to rewrite the input queries to be answered using those cubes. In Kylin, query plans are executed using a combination of Calcite native operators and HBase.

Recently Calcite has become popular among streaming systems too. Projects such as Apache Apex~\cite{website:Apex},  Flink~\cite{website:Flink}, Apache Samza~\cite{website:Samza}, and  Storm~\cite{website:Storm} have chosen to integrate with Calcite, using its components to provide a streaming SQL interface to their users. Finally, other commercial systems have adopted Calcite, such as MapD~\cite{website:MapD}, Lingual~\cite{website:Lingual}, and Qubole Quark~\cite{website:Qubole}.

{%
\renewcommand{\arraystretch}{1.5}
\begin{table}[t]
\centering
{\begin{tabular}{cc} \toprule
\textbf{Adapter} & \textbf{Target language} \\
\midrule
Apache Cassandra & Cassandra Query Language (CQL) \\
Apache Pig & Pig Latin \\
Apache Spark & Java (Resilient Distributed Datasets) \\
Druid & JSON \\
Elasticsearch & JSON \\
JDBC & SQL (multiple dialects) \\
MongoDB & Java \\
Splunk & SPL \\
\bottomrule
\end{tabular}}
\caption{List of Calcite adapters.\label{tab:adapters}}\vspaceafigure\vspaceafigure
\end{table}
}


\subsection{Calcite Adapters}

Instead of using Calcite as a library, other systems integrate with Calcite via adapters which read their data sources.
Table~\ref{tab:adapters} provides the list of available adapters in Calcite.
One of the main key components of the implementation of these adapters is the \textit{converter} responsible for translating the algebra expression to be pushed to the system into the query language supported by that system.
Table~\ref{tab:adapters} also shows the languages that Calcite translates into for each of these adapters.

The JDBC adapter supports the generation of multiple SQL dialects, including those supported by popular RDBMSes such as PostgreSQL and MySQL\@.
In turn, the adapter for Cassandra~\cite{website:Cassandra} generates its own SQL-like language called \textit{CQL} whereas the adapter for Apache Pig~\cite{website:Pig} generates queries expressed in Pig Latin~\cite{Olston08a}.
The adapter for Apache Spark~\cite{website:Spark} uses the Java RDD API\@.
Finally, Druid~\cite{website:Druid}, Elasticsearch~\cite{website:Elastic} and Splunk~\cite{website:Splunk} are queried through REST HTTP API requests.
The queries generated by Calcite for these systems are expressed in JSON or XML.

\subsection{Uses in Research}

In a research setting, Calcite has been considered~\cite{calcite2017polystore} as a polystore-alternative for precision medicine and clinical analysis scenarios.
In those scenarios, heterogeneous medical data has to be logically assembled and aligned  to assess the best treatments based on the comprehensive medical history and the genomic profile of the patient. The data comes from relational sources representing patients' electronic medical records, structured and semi-structured sources representing various reports (oncology, psychiatry,laboratory tests, radiology, etc.), imaging, signals, and sequence data, stored in scientific databases.
In those circumstances, Calcite represents a good foundation with its uniform query interface, and flexible adapter architecture, but the ongoing research efforts are aimed at ($i$)~introduction of the new adapters for array, and textual sources, and ($ii$)~support efficient joining of heterogeneous data sources.

\section{Future Work}
\label{sec:future}

The future work on Calcite will focus on the development of the new features, and the expansion of its adapter architecture:

\begin{itemize}
\item Enhancements to the design of Calcite to further support its use a standalone
engine, which would require a support for data definition languages (DDL), materialized views, indexes and constraints.
\item Ongoing improvements to the design and flexibility of the planner, including making it more modular, allowing users Calcite to supply planner \textit{programs} (collections of rules organized into
planning phases) for execution.
\item Incorporation of new parametric approaches~\cite{trummer2017multi} into the design of the optimizer.
\item Support for an extended set of SQL commands, functions, and utilities, including full compliance with OpenGIS.
\item New adapters for non-relational data sources such as array databases for scientific computing.
\item Improvements to performance profiling and instrumentation.
\end{itemize}

\subsection{Performance Testing and Evaluation}
\label{sec:perf}
Though Calcite contains a performance testing module, it does not evaluate query execution.
It would  be useful to assess the performance of systems built with Calcite.
For example, we could compare the performance of Calcite with similar frameworks.
Unfortunately, it might be difficult to craft fair comparisons.
For example, like Calcite, Algebricks optimizes queries for Hive.
Borkar et al.~\cite{Borkar:2015:ADM:2806777.2806941} compared
Algebricks with the Hyracks scheduler against Hive~version~0.12 (without Calcite). The work of Borkar et al.\ precedes significant engineering and architectural changes into Hive.
Comparing Calcite against Algebricks in a fair manner in terms of timings does not seem feasible, as one would need to ensure that each uses the same execution engine.
Hive applications rely mostly on either Apache Tez or Apache Spark as execution engines whereas Algebricks is tied to its own framework (including Hyracks).

Moreover, to assess the performance of Calcite-based systems, we need to consider two distinct use cases. Indeed, Calcite can be used either as part of a single system---as a tool to accelerate the construction of such a system---or for the more difficult task of combining several distinct systems---as  a common layer.
The former is tied to the characteristics of the data processing system, and because Calcite is so versatile and widely used, many distinct benchmarks are needed. The latter is limited by the availability of existing heterogeneous benchmarks. BigDAWG~\cite{8091081} has been used to integrate PostgreSQL with Vertica, and on a standard benchmark, one gets that the integrated system is superior to a baseline where entire tables are copied from one system to another to answer specific queries. Based on real-world experience, we believe that more ambitious goals are possible for integrated multiple systems: they should be superior to the sum of their parts.

\section{Conclusion}
\label{sec:conclusion}
Emerging data management practices and associated analytic uses of data continue to evolve towards an increasingly diverse, and heterogeneous spectrum of scenarios.
At the same time, relational data sources, accessed through the SQL, remain an essential means to how enterprises work with the data.
In this somewhat dichotomous space,  Calcite plays a unique role with its strong support for both traditional, conventional data processing, and for its support of other data sources including those with semi-structured, streaming and geospatial models.
In addition, Calcite's design philosophy with a focus on flexibility, adaptivity, and extensibility, has been another factor in Calcite becoming the most widely adopted query optimizer, used in a large number of open-source frameworks.
Calcite's dynamic and flexible query optimizer, and its adapter architecture allows it to be embedded selectively by a variety of data management frameworks such as Hive, Drill, MapD, and Flink.
Calcite's support for heterogeneous data processing, as well as for the extended set of relational functions will continue to improve, in both functionality and performance.

\section*{Acknowledgments}
We would like to thank the  Calcite community, contributors and users, who build, maintain, use, test, write about, and continue to push the Calcite project forward.
This manuscript has been in part co-authored by UT-Battelle, LLC under Contract No. DE-AC05-00OR22725 with the U.S. Department of Energy.

\bibliographystyle{ACM-Reference-Format}
\bibliography{references}


\begin{thebibliography}{57}


\ifx \showCODEN    \undefined \def \showCODEN     #1{\unskip}     \fi
\ifx \showDOI      \undefined \def \showDOI       #1{#1}\fi
\ifx \showISBNx    \undefined \def \showISBNx     #1{\unskip}     \fi
\ifx \showISBNxiii \undefined \def \showISBNxiii  #1{\unskip}     \fi
\ifx \showISSN     \undefined \def \showISSN      #1{\unskip}     \fi
\ifx \showLCCN     \undefined \def \showLCCN      #1{\unskip}     \fi
\ifx \shownote     \undefined \def \shownote      #1{#1}          \fi
\ifx \showarticletitle \undefined \def \showarticletitle #1{#1}   \fi
\ifx \showURL      \undefined \def \showURL       {\relax}        \fi
\providecommand\bibfield[2]{#2}
\providecommand\bibinfo[2]{#2}
\providecommand\natexlab[1]{#1}
\providecommand\showeprint[2][]{arXiv:#2}

\bibitem[\protect\citeauthoryear{Apex}{Apex}{2017}]%
        {website:Apex}
Apex.
\newblock \bibinfo{title}{{Apache Apex}}.
\newblock \bibinfo{howpublished}{\url{https://apex.apache.org}}.
  (\bibinfo{date}{Nov.} \bibinfo{year}{2017}).
\newblock


\bibitem[\protect\citeauthoryear{Arasu, Babu, and Widom}{Arasu
  et~al\mbox{.}}{2003}]%
        {ilprints758}
\bibfield{author}{\bibinfo{person}{Arvind Arasu}, \bibinfo{person}{Shivnath
  Babu}, {and} \bibinfo{person}{Jennifer Widom}.}
  \bibinfo{year}{2003}\natexlab{}.
\newblock \bibinfo{booktitle}{{\em The {CQL} Continuous Query Language:
  Semantic Foundations and Query Execution}}.
\newblock \bibinfo{type}{Technical Report} 2003-67.
  \bibinfo{institution}{Stanford InfoLab}.
\newblock


\bibitem[\protect\citeauthoryear{Armbrust et~al\mbox{.}}{Armbrust
  et~al\mbox{.}}{2015a}]%
        {Armbrust2015}
\bibfield{author}{\bibinfo{person}{Michael Armbrust} {et~al\mbox{.}}}
  \bibinfo{year}{2015}\natexlab{a}.
\newblock \showarticletitle{Spark {SQL}: Relational Data Processing in
  {Spark}}. In \bibinfo{booktitle}{{\em Proceedings of the 2015 ACM SIGMOD
  International Conference on Management of Data}} {\em
  (\bibinfo{series}{SIGMOD '15})}. \bibinfo{publisher}{ACM},
  \bibinfo{address}{New York, NY, USA}, \bibinfo{pages}{1383--1394}.
\newblock


\bibitem[\protect\citeauthoryear{Armbrust, Xin, Lian, Huai, Liu, Bradley, Meng,
  Kaftan, Franklin, Ghodsi, and Zaharia}{Armbrust et~al\mbox{.}}{2015b}]%
        {DBLP:conf/sigmod/ArmbrustXLHLBMK15}
\bibfield{author}{\bibinfo{person}{Michael Armbrust},
  \bibinfo{person}{Reynold~S. Xin}, \bibinfo{person}{Cheng Lian},
  \bibinfo{person}{Yin Huai}, \bibinfo{person}{Davies Liu},
  \bibinfo{person}{Joseph~K. Bradley}, \bibinfo{person}{Xiangrui Meng},
  \bibinfo{person}{Tomer Kaftan}, \bibinfo{person}{Michael~J. Franklin},
  \bibinfo{person}{Ali Ghodsi}, {and} \bibinfo{person}{Matei Zaharia}.}
  \bibinfo{year}{2015}\natexlab{b}.
\newblock \showarticletitle{{Spark SQL: Relational Data Processing in Spark}}.
  In \bibinfo{booktitle}{{\em Proceedings of the 2015 ACM SIGMOD International
  Conference on Management of Data}} {\em (\bibinfo{series}{SIGMOD '15})}.
  \bibinfo{publisher}{ACM}, \bibinfo{address}{New York, NY, USA},
  \bibinfo{pages}{1383--1394}.
\newblock
\showISBNx{978-1-4503-2758-9}


\bibitem[\protect\citeauthoryear{ASF}{ASF}{2017}]%
        {asf:website}
\bibfield{author}{\bibinfo{person}{ASF}.}
\newblock \bibinfo{title}{{The Apache Software Foundation}}.
\newblock   (\bibinfo{date}{Nov.} \bibinfo{year}{2017}).
\newblock
\showURL{%
Retrieved November 20, 2017 from \url{http://www.apache.org/}}


\bibitem[\protect\citeauthoryear{Borkar, Bu, Carman, Onose, Westmann, Pirzadeh,
  Carey, and Tsotras}{Borkar et~al\mbox{.}}{2015}]%
        {Borkar:2015:ADM:2806777.2806941}
\bibfield{author}{\bibinfo{person}{Vinayak Borkar}, \bibinfo{person}{Yingyi
  Bu}, \bibinfo{person}{E.~Preston Carman, Jr.}, \bibinfo{person}{Nicola
  Onose}, \bibinfo{person}{Till Westmann}, \bibinfo{person}{Pouria Pirzadeh},
  \bibinfo{person}{Michael~J. Carey}, {and} \bibinfo{person}{Vassilis~J.
  Tsotras}.} \bibinfo{year}{2015}\natexlab{}.
\newblock \showarticletitle{Algebricks: A Data Model-agnostic Compiler Backend
  for Big Data Languages}. In \bibinfo{booktitle}{{\em Proceedings of the Sixth
  ACM Symposium on Cloud Computing}} {\em (\bibinfo{series}{SoCC '15})}.
  \bibinfo{publisher}{ACM}, \bibinfo{address}{New York, NY, USA},
  \bibinfo{pages}{422--433}.
\newblock
\showISBNx{978-1-4503-3651-2}


\bibitem[\protect\citeauthoryear{Carey et~al\mbox{.}}{Carey
  et~al\mbox{.}}{1995}]%
        {Carey1995}
\bibfield{author}{\bibinfo{person}{M.~J. Carey} {et~al\mbox{.}}}
  \bibinfo{year}{1995}\natexlab{}.
\newblock \showarticletitle{Towards heterogeneous multimedia information
  systems: the {Garlic} approach}. In \bibinfo{booktitle}{{\em IDE-DOM '95}}.
  \bibinfo{pages}{124--131}.
\newblock


\bibitem[\protect\citeauthoryear{Cassandra}{Cassandra}{2017}]%
        {website:Cassandra}
Cassandra.
\newblock \bibinfo{title}{{Apache Cassandra}}.
\newblock   (\bibinfo{date}{Nov.} \bibinfo{year}{2017}).
\newblock
\showURL{%
Retrieved November 20, 2017 from \url{http://cassandra.apache.org/}}


\bibitem[\protect\citeauthoryear{Chang, Dean, Ghemawat, Hsieh, Wallach,
  Burrows, Chandra, Fikes, and Gruber}{Chang et~al\mbox{.}}{2006}]%
        {DBLP:conf/osdi/ChangDGHWBCFG06}
\bibfield{author}{\bibinfo{person}{Fay Chang}, \bibinfo{person}{Jeffrey Dean},
  \bibinfo{person}{Sanjay Ghemawat}, \bibinfo{person}{Wilson~C. Hsieh},
  \bibinfo{person}{Deborah~A. Wallach}, \bibinfo{person}{Michael Burrows},
  \bibinfo{person}{Tushar Chandra}, \bibinfo{person}{Andrew Fikes}, {and}
  \bibinfo{person}{Robert Gruber}.} \bibinfo{year}{2006}\natexlab{}.
\newblock \showarticletitle{{Bigtable: {A} Distributed Storage System for
  Structured Data}}. In \bibinfo{booktitle}{{\em 7th Symposium on Operating
  Systems Design and Implementation {(OSDI} '06), November 6-8, Seattle, WA,
  {USA}}}. \bibinfo{pages}{205--218}.
\newblock


\bibitem[\protect\citeauthoryear{Chaudhuri, Krishnamurthy, Potamianos, and
  Shim}{Chaudhuri et~al\mbox{.}}{1995}]%
        {DBLP:conf/icde/ChaudhuriKPS95}
\bibfield{author}{\bibinfo{person}{Surajit Chaudhuri}, \bibinfo{person}{Ravi
  Krishnamurthy}, \bibinfo{person}{Spyros Potamianos}, {and}
  \bibinfo{person}{Kyuseok Shim}.} \bibinfo{year}{1995}\natexlab{}.
\newblock \showarticletitle{{Optimizing Queries with Materialized Views}}. In
  \bibinfo{booktitle}{{\em Proceedings of the Eleventh International Conference
  on Data Engineering}} {\em (\bibinfo{series}{ICDE '95})}.
  \bibinfo{publisher}{IEEE Computer Society}, \bibinfo{address}{Washington, DC,
  USA}, \bibinfo{pages}{190--200}.
\newblock
\showISBNx{0-8186-6910-1}


\bibitem[\protect\citeauthoryear{Codd}{Codd}{1970}]%
        {DBLP:journals/cacm/Codd70}
\bibfield{author}{\bibinfo{person}{E.~F. Codd}.}
  \bibinfo{year}{1970}\natexlab{}.
\newblock \showarticletitle{A Relational Model of Data for Large Shared Data
  Banks}.
\newblock \bibinfo{journal}{{\em Commun. ACM\/}} \bibinfo{volume}{13},
  \bibinfo{number}{6} (\bibinfo{date}{June} \bibinfo{year}{1970}),
  \bibinfo{pages}{377--387}.
\newblock
\showISSN{0001-0782}


\bibitem[\protect\citeauthoryear{\c{S}uhan}{\c{S}uhan}{2017}]%
        {website:MapDblog}
\bibfield{author}{\bibinfo{person}{Alex \c{S}uhan}.}
\newblock \bibinfo{title}{{Fast and Flexible Query Analysis at MapD with Apache
  Calcite}}.
\newblock   (\bibinfo{date}{feb} \bibinfo{year}{2017}).
\newblock
\showURL{%
Retrieved November 20, 2017 from
  \url{https://www.mapd.com/blog/2017/02/08/fast-and-flexible-query-analysis-at-mapd-with-apache-calcite-2/}}


\bibitem[\protect\citeauthoryear{Drill}{Drill}{2017}]%
        {website:Drill}
Drill.
\newblock \bibinfo{title}{{Apache Drill}}.
\newblock   (\bibinfo{date}{Nov.} \bibinfo{year}{2017}).
\newblock
\showURL{%
Retrieved November 20, 2017 from \url{http://drill.apache.org/}}


\bibitem[\protect\citeauthoryear{Druid}{Druid}{2017}]%
        {website:Druid}
Druid.
\newblock \bibinfo{title}{{Druid}}.
\newblock   (\bibinfo{date}{Nov.} \bibinfo{year}{2017}).
\newblock
\showURL{%
Retrieved November 20, 2017 from \url{http://druid.io/}}


\bibitem[\protect\citeauthoryear{Elastic}{Elastic}{2017}]%
        {website:Elastic}
Elastic.
\newblock \bibinfo{title}{{Elasticsearch}}.
\newblock   (\bibinfo{date}{Nov.} \bibinfo{year}{2017}).
\newblock
\showURL{%
Retrieved November 20, 2017 from \url{https://www.elastic.co}}


\bibitem[\protect\citeauthoryear{Flink}{Flink}{2017}]%
        {website:Flink}
Flink.
\newblock \bibinfo{title}{{Apache Flink}}.
\newblock \bibinfo{howpublished}{\url{https://flink.apache.org}}.
  (\bibinfo{date}{Nov.} \bibinfo{year}{2017}).
\newblock


\bibitem[\protect\citeauthoryear{Fu, Ong, Papakonstantinou, and Petropoulos}{Fu
  et~al\mbox{.}}{2011}]%
        {fu2011sql}
\bibfield{author}{\bibinfo{person}{Yupeng Fu}, \bibinfo{person}{Kian~Win Ong},
  \bibinfo{person}{Yannis Papakonstantinou}, {and} \bibinfo{person}{Michalis
  Petropoulos}.} \bibinfo{year}{2011}\natexlab{}.
\newblock \showarticletitle{{The SQL-based all-declarative FORWARD web
  application development framework}}. In \bibinfo{booktitle}{{\em CIDR}}.
\newblock


\bibitem[\protect\citeauthoryear{Goldstein and Larson}{Goldstein and
  Larson}{2001}]%
        {DBLP:conf/sigmod/GoldsteinL01}
\bibfield{author}{\bibinfo{person}{Jonathan Goldstein} {and}
  \bibinfo{person}{Per-\r{A}ke Larson}.} \bibinfo{year}{2001}\natexlab{}.
\newblock \showarticletitle{Optimizing Queries Using Materialized Views: A
  Practical, Scalable Solution}.
\newblock \bibinfo{journal}{{\em SIGMOD Rec.\/}} \bibinfo{volume}{30},
  \bibinfo{number}{2} (\bibinfo{date}{May} \bibinfo{year}{2001}),
  \bibinfo{pages}{331--342}.
\newblock
\showISSN{0163-5808}


\bibitem[\protect\citeauthoryear{Graefe}{Graefe}{1995}]%
        {DBLP:journals/debu/Graefe95a}
\bibfield{author}{\bibinfo{person}{Goetz Graefe}.}
  \bibinfo{year}{1995}\natexlab{}.
\newblock \showarticletitle{{The Cascades Framework for Query Optimization}}.
\newblock \bibinfo{journal}{{\em {IEEE} Data Eng. Bull.\/}}
  (\bibinfo{year}{1995}).
\newblock


\bibitem[\protect\citeauthoryear{Graefe and McKenna}{Graefe and
  McKenna}{1993}]%
        {Graefe93thevolcano}
\bibfield{author}{\bibinfo{person}{Goetz Graefe} {and}
  \bibinfo{person}{William~J. McKenna}.} \bibinfo{year}{1993}\natexlab{}.
\newblock \showarticletitle{{The Volcano Optimizer Generator: Extensibility and
  Efficient Search}}. In \bibinfo{booktitle}{{\em Proceedings of the Ninth
  International Conference on Data Engineering}}. \bibinfo{publisher}{IEEE
  Computer Society}, \bibinfo{address}{Washington, DC, USA},
  \bibinfo{pages}{209--218}.
\newblock
\showISBNx{0-8186-3570-3}


\bibitem[\protect\citeauthoryear{Halperin, Teixeira~de Almeida, Choo, Chu,
  Koutris, Moritz, Ortiz, Ruamviboonsuk, Wang, Whitaker, Xu, Balazinska, Howe,
  and Suciu}{Halperin et~al\mbox{.}}{2014}]%
        {DBLP:conf/sigmod/HalperinACCKMORWWXBHS14}
\bibfield{author}{\bibinfo{person}{Daniel Halperin}, \bibinfo{person}{Victor
  Teixeira~de Almeida}, \bibinfo{person}{Lee~Lee Choo}, \bibinfo{person}{Shumo
  Chu}, \bibinfo{person}{Paraschos Koutris}, \bibinfo{person}{Dominik Moritz},
  \bibinfo{person}{Jennifer Ortiz}, \bibinfo{person}{Vaspol Ruamviboonsuk},
  \bibinfo{person}{Jingjing Wang}, \bibinfo{person}{Andrew Whitaker},
  \bibinfo{person}{Shengliang Xu}, \bibinfo{person}{Magdalena Balazinska},
  \bibinfo{person}{Bill Howe}, {and} \bibinfo{person}{Dan Suciu}.}
  \bibinfo{year}{2014}\natexlab{}.
\newblock \showarticletitle{Demonstration of the Myria Big Data Management
  Service}. In \bibinfo{booktitle}{{\em Proceedings of the 2014 ACM SIGMOD
  International Conference on Management of Data}} {\em
  (\bibinfo{series}{SIGMOD '14})}. \bibinfo{publisher}{ACM},
  \bibinfo{address}{New York, NY, USA}, \bibinfo{pages}{881--884}.
\newblock
\showISBNx{978-1-4503-2376-5}


\bibitem[\protect\citeauthoryear{Harinarayan, Rajaraman, and
  Ullman}{Harinarayan et~al\mbox{.}}{1996}]%
        {DataCubes}
\bibfield{author}{\bibinfo{person}{Venky Harinarayan}, \bibinfo{person}{Anand
  Rajaraman}, {and} \bibinfo{person}{Jeffrey~D. Ullman}.}
  \bibinfo{year}{1996}\natexlab{}.
\newblock \showarticletitle{{Implementing Data Cubes Efficiently}}.
\newblock \bibinfo{journal}{{\em SIGMOD Rec.\/}} \bibinfo{volume}{25},
  \bibinfo{number}{2} (\bibinfo{date}{June} \bibinfo{year}{1996}),
  \bibinfo{pages}{205--216}.
\newblock
\showISSN{0163-5808}


\bibitem[\protect\citeauthoryear{HBase}{HBase}{2017}]%
        {website:HBase}
HBase.
\newblock \bibinfo{title}{{Apache HBase}}.
\newblock   (\bibinfo{date}{Nov.} \bibinfo{year}{2017}).
\newblock
\showURL{%
Retrieved November 20, 2017 from \url{http://hbase.apache.org/}}


\bibitem[\protect\citeauthoryear{Hive}{Hive}{2017}]%
        {website:Hive}
Hive.
\newblock \bibinfo{title}{{Apache Hive}}.
\newblock   (\bibinfo{date}{Nov.} \bibinfo{year}{2017}).
\newblock
\showURL{%
Retrieved November 20, 2017 from \url{http://hive.apache.org/}}


\bibitem[\protect\citeauthoryear{Huai, Chauhan, Gates, Hagleitner, Hanson,
  O'Malley, Pandey, Yuan, Lee, and Zhang}{Huai et~al\mbox{.}}{2014}]%
        {DBLP:conf/sigmod/HuaiCGHHOPYL014}
\bibfield{author}{\bibinfo{person}{Yin Huai}, \bibinfo{person}{Ashutosh
  Chauhan}, \bibinfo{person}{Alan Gates}, \bibinfo{person}{Gunther Hagleitner},
  \bibinfo{person}{Eric~N. Hanson}, \bibinfo{person}{Owen O'Malley},
  \bibinfo{person}{Jitendra Pandey}, \bibinfo{person}{Yuan Yuan},
  \bibinfo{person}{Rubao Lee}, {and} \bibinfo{person}{Xiaodong Zhang}.}
  \bibinfo{year}{2014}\natexlab{}.
\newblock \showarticletitle{{Major Technical Advancements in Apache Hive}}. In
  \bibinfo{booktitle}{{\em Proceedings of the 2014 ACM SIGMOD International
  Conference on Management of Data}} {\em (\bibinfo{series}{SIGMOD '14})}.
  \bibinfo{publisher}{ACM}, \bibinfo{address}{New York, NY, USA},
  \bibinfo{pages}{1235--1246}.
\newblock
\showISBNx{978-1-4503-2376-5}


\bibitem[\protect\citeauthoryear{Hyde}{Hyde}{2010}]%
        {DBLP:journals/cacm/Hyde10}
\bibfield{author}{\bibinfo{person}{Julian Hyde}.}
  \bibinfo{year}{2010}\natexlab{}.
\newblock \showarticletitle{{Data in Flight}}.
\newblock \bibinfo{journal}{{\em Commun. ACM\/}} \bibinfo{volume}{53},
  \bibinfo{number}{1} (\bibinfo{date}{Jan.} \bibinfo{year}{2010}),
  \bibinfo{pages}{48--52}.
\newblock
\showISSN{0001-0782}


\bibitem[\protect\citeauthoryear{Janino}{Janino}{2017}]%
        {website:Janino}
Janino.
\newblock \bibinfo{title}{{Janino: A super-small, super-fast Java compiler}}.
\newblock   (\bibinfo{date}{Nov.} \bibinfo{year}{2017}).
\newblock
\showURL{%
Retrieved November 20, 2017 from \url{http://www.janino.net/}}


\bibitem[\protect\citeauthoryear{Kylin}{Kylin}{2017}]%
        {website:Kylin}
Kylin.
\newblock \bibinfo{title}{{Apache Kylin}}.
\newblock   (\bibinfo{date}{Nov.} \bibinfo{year}{2017}).
\newblock
\showURL{%
Retrieved November 20, 2017 from \url{http://kylin.apache.org/}}


\bibitem[\protect\citeauthoryear{Lakshman and Malik}{Lakshman and
  Malik}{2010}]%
        {Cassandra}
\bibfield{author}{\bibinfo{person}{Avinash Lakshman} {and}
  \bibinfo{person}{Prashant Malik}.} \bibinfo{year}{2010}\natexlab{}.
\newblock \showarticletitle{{Cassandra: A Decentralized Structured Storage
  System}}.
\newblock \bibinfo{journal}{{\em SIGOPS Oper. Syst. Rev.\/}}
  \bibinfo{volume}{44}, \bibinfo{number}{2} (\bibinfo{date}{April}
  \bibinfo{year}{2010}), \bibinfo{pages}{35--40}.
\newblock
\showISSN{0163-5980}


\bibitem[\protect\citeauthoryear{Lingual}{Lingual}{2017}]%
        {website:Lingual}
Lingual.
\newblock \bibinfo{title}{{Lingual}}.
\newblock   (\bibinfo{date}{Nov.} \bibinfo{year}{2017}).
\newblock
\showURL{%
Retrieved November 20, 2017 from
  \url{http://www.cascading.org/projects/lingual/}}


\bibitem[\protect\citeauthoryear{Lucene}{Lucene}{2017}]%
        {website:Lucene}
Lucene.
\newblock \bibinfo{title}{{Apache Lucene}}.
\newblock   (\bibinfo{date}{Nov.} \bibinfo{year}{2017}).
\newblock
\showURL{%
Retrieved November 20, 2017 from \url{https://lucene.apache.org/}}


\bibitem[\protect\citeauthoryear{MapD}{MapD}{2017}]%
        {website:MapD}
MapD.
\newblock \bibinfo{title}{{MapD}}.
\newblock   (\bibinfo{date}{Nov.} \bibinfo{year}{2017}).
\newblock
\showURL{%
Retrieved November 20, 2017 from \url{https://www.mapd.com}}


\bibitem[\protect\citeauthoryear{Meijer, Beckman, and Bierman}{Meijer
  et~al\mbox{.}}{2006}]%
        {DBLP:conf/sigmod/MeijerBB06}
\bibfield{author}{\bibinfo{person}{Erik Meijer}, \bibinfo{person}{Brian
  Beckman}, {and} \bibinfo{person}{Gavin Bierman}.}
  \bibinfo{year}{2006}\natexlab{}.
\newblock \showarticletitle{LINQ: Reconciling Object, Relations and XML in the
  .NET Framework}. In \bibinfo{booktitle}{{\em Proceedings of the 2006 ACM
  SIGMOD International Conference on Management of Data}} {\em
  (\bibinfo{series}{SIGMOD '06})}. \bibinfo{publisher}{ACM},
  \bibinfo{address}{New York, NY, USA}, \bibinfo{pages}{706--706}.
\newblock
\showISBNx{1-59593-434-0}


\bibitem[\protect\citeauthoryear{Melnik, Gubarev, Long, Romer, Shivakumar,
  Tolton, and Vassilakis}{Melnik et~al\mbox{.}}{2010}]%
        {DBLP:journals/pvldb/MelnikGLRSTV10}
\bibfield{author}{\bibinfo{person}{Sergey Melnik}, \bibinfo{person}{Andrey
  Gubarev}, \bibinfo{person}{Jing~Jing Long}, \bibinfo{person}{Geoffrey Romer},
  \bibinfo{person}{Shiva Shivakumar}, \bibinfo{person}{Matt Tolton}, {and}
  \bibinfo{person}{Theo Vassilakis}.} \bibinfo{year}{2010}\natexlab{}.
\newblock \showarticletitle{{Dremel: Interactive Analysis of Web-Scale
  Datasets}}.
\newblock \bibinfo{journal}{{\em {PVLDB}\/}} \bibinfo{volume}{3},
  \bibinfo{number}{1} (\bibinfo{year}{2010}), \bibinfo{pages}{330--339}.
\newblock
\showURL{%
\url{http://www.comp.nus.edu.sg/~vldb2010/proceedings/files/papers/R29.pdf}}


\bibitem[\protect\citeauthoryear{Mendes, Bizarro, and Marques}{Mendes
  et~al\mbox{.}}{2009}]%
        {mendes2009performance}
\bibfield{author}{\bibinfo{person}{Marcelo~RN Mendes}, \bibinfo{person}{Pedro
  Bizarro}, {and} \bibinfo{person}{Paulo Marques}.}
  \bibinfo{year}{2009}\natexlab{}.
\newblock \showarticletitle{A performance study of event processing systems}.
  In \bibinfo{booktitle}{{\em Technology Conference on Performance Evaluation
  and Benchmarking}}. Springer, \bibinfo{pages}{221--236}.
\newblock


\bibitem[\protect\citeauthoryear{Mongo}{Mongo}{2017}]%
        {website:Mongo}
Mongo.
\newblock \bibinfo{title}{{MongoDB}}.
\newblock   (\bibinfo{date}{Nov.} \bibinfo{year}{2017}).
\newblock
\showURL{%
Retrieved November 28, 2017 from \url{https://www.mongodb.com/}}


\bibitem[\protect\citeauthoryear{Olston, Reed, Srivastava, Kumar, and
  Tomkins}{Olston et~al\mbox{.}}{2008}]%
        {Olston08a}
\bibfield{author}{\bibinfo{person}{Christopher Olston},
  \bibinfo{person}{Benjamin Reed}, \bibinfo{person}{Utkarsh Srivastava},
  \bibinfo{person}{Ravi Kumar}, {and} \bibinfo{person}{Andrew Tomkins}.}
  \bibinfo{year}{2008}\natexlab{}.
\newblock \showarticletitle{{Pig Latin: a not-so-foreign language for data
  processing}}. In \bibinfo{booktitle}{{\em SIGMOD}}.
\newblock


\bibitem[\protect\citeauthoryear{Ong, Papakonstantinou, and Vernoux}{Ong
  et~al\mbox{.}}{2014}]%
        {ong2014sql++}
\bibfield{author}{\bibinfo{person}{Kian~Win Ong}, \bibinfo{person}{Yannis
  Papakonstantinou}, {and} \bibinfo{person}{Romain Vernoux}.}
  \bibinfo{year}{2014}\natexlab{}.
\newblock \showarticletitle{{The SQL++ query language: Configurable, unifying
  and semi-structured}}.
\newblock \bibinfo{journal}{{\em arXiv preprint arXiv:1405.3631\/}}
  (\bibinfo{year}{2014}).
\newblock


\bibitem[\protect\citeauthoryear{Open Geospatial Consortium}{Open Geospatial
  Consortium}{2010}]%
        {OpenGIS:SFA}
Open Geospatial Consortium.
\newblock \bibinfo{title}{{OpenGIS Implementation Specification for Geographic
  information - Simple feature access - Part 2: SQL option}}.
\newblock
  \bibinfo{howpublished}{\url{http://portal.opengeospatial.org/files/?artifact_id=25355}}.
    (\bibinfo{year}{2010}).
\newblock


\bibitem[\protect\citeauthoryear{Phoenix}{Phoenix}{2017}]%
        {website:Phoenix}
Phoenix.
\newblock \bibinfo{title}{{Apache Phoenix}}.
\newblock   (\bibinfo{date}{Nov.} \bibinfo{year}{2017}).
\newblock
\showURL{%
Retrieved November 20, 2017 from \url{http://phoenix.apache.org/}}


\bibitem[\protect\citeauthoryear{Pig}{Pig}{2017}]%
        {website:Pig}
Pig.
\newblock \bibinfo{title}{{Apache Pig}}.
\newblock   (\bibinfo{date}{Nov.} \bibinfo{year}{2017}).
\newblock
\showURL{%
Retrieved November 20, 2017 from \url{http://pig.apache.org/}}


\bibitem[\protect\citeauthoryear{Qubole Quark}{Qubole Quark}{2017}]%
        {website:Qubole}
Qubole Quark.
\newblock \bibinfo{title}{{Qubole Quark}}.
\newblock   (\bibinfo{date}{Nov.} \bibinfo{year}{2017}).
\newblock
\showURL{%
Retrieved November 20, 2017 from \url{https://github.com/qubole/quark}}


\bibitem[\protect\citeauthoryear{Saha, Shah, Seth, Vijayaraghavan, Murthy, and
  Curino}{Saha et~al\mbox{.}}{2015}]%
        {DBLP:conf/sigmod/SahaSSVMC15}
\bibfield{author}{\bibinfo{person}{Bikas Saha}, \bibinfo{person}{Hitesh Shah},
  \bibinfo{person}{Siddharth Seth}, \bibinfo{person}{Gopal Vijayaraghavan},
  \bibinfo{person}{Arun~C. Murthy}, {and} \bibinfo{person}{Carlo Curino}.}
  \bibinfo{year}{2015}\natexlab{}.
\newblock \showarticletitle{Apache Tez: {A} Unifying Framework for Modeling and
  Building Data Processing Applications}. In \bibinfo{booktitle}{{\em
  Proceedings of the 2015 {ACM} {SIGMOD} International Conference on Management
  of Data, Melbourne, Victoria, Australia, May 31 - June 4, 2015}}.
  \bibinfo{pages}{1357--1369}.
\newblock
\showDOI{%
\url{https://doi.org/10.1145/2723372.2742790}}


\bibitem[\protect\citeauthoryear{Samza}{Samza}{2017}]%
        {website:Samza}
Samza.
\newblock \bibinfo{title}{{Apache Samza}}.
\newblock   (\bibinfo{date}{Nov.} \bibinfo{year}{2017}).
\newblock
\showURL{%
Retrieved November 20, 2017 from \url{http://samza.apache.org/}}


\bibitem[\protect\citeauthoryear{Soliman, Antova, Raghavan, El-Helw, Gu, Shen,
  Caragea, Garcia-Alvarado, Rahman, Petropoulos, Waas, Narayanan, Krikellas,
  and Baldwin}{Soliman et~al\mbox{.}}{2014}]%
        {Soliman:2014:OMQ:2588555.2595637}
\bibfield{author}{\bibinfo{person}{Mohamed~A. Soliman},
  \bibinfo{person}{Lyublena Antova}, \bibinfo{person}{Venkatesh Raghavan},
  \bibinfo{person}{Amr El-Helw}, \bibinfo{person}{Zhongxian Gu},
  \bibinfo{person}{Entong Shen}, \bibinfo{person}{George~C. Caragea},
  \bibinfo{person}{Carlos Garcia-Alvarado}, \bibinfo{person}{Foyzur Rahman},
  \bibinfo{person}{Michalis Petropoulos}, \bibinfo{person}{Florian Waas},
  \bibinfo{person}{Sivaramakrishnan Narayanan}, \bibinfo{person}{Konstantinos
  Krikellas}, {and} \bibinfo{person}{Rhonda Baldwin}.}
  \bibinfo{year}{2014}\natexlab{}.
\newblock \showarticletitle{{Orca: A Modular Query Optimizer Architecture for
  Big Data}}. In \bibinfo{booktitle}{{\em Proceedings of the 2014 ACM SIGMOD
  International Conference on Management of Data}} {\em
  (\bibinfo{series}{SIGMOD '14})}. \bibinfo{publisher}{ACM},
  \bibinfo{address}{New York, NY, USA}, \bibinfo{pages}{337--348}.
\newblock
\showISBNx{978-1-4503-2376-5}


\bibitem[\protect\citeauthoryear{Solr}{Solr}{2017}]%
        {website:Solr}
Solr.
\newblock \bibinfo{title}{{Apache Solr}}.
\newblock   (\bibinfo{date}{Nov.} \bibinfo{year}{2017}).
\newblock
\showURL{%
Retrieved November 20, 2017 from \url{http://lucene.apache.org/solr/}}


\bibitem[\protect\citeauthoryear{Spark}{Spark}{2017}]%
        {website:Spark}
Spark.
\newblock \bibinfo{title}{{Apache Spark}}.
\newblock   (\bibinfo{date}{Nov.} \bibinfo{year}{2017}).
\newblock
\showURL{%
Retrieved November 20, 2017 from \url{http://spark.apache.org/}}


\bibitem[\protect\citeauthoryear{Splunk}{Splunk}{2017}]%
        {website:Splunk}
Splunk.
\newblock \bibinfo{title}{{Splunk}}.
\newblock   (\bibinfo{date}{Nov.} \bibinfo{year}{2017}).
\newblock
\showURL{%
Retrieved November 20, 2017 from \url{https://www.splunk.com/}}


\bibitem[\protect\citeauthoryear{Stonebraker and {\c{C}}etintemel}{Stonebraker
  and {\c{C}}etintemel}{2005}]%
        {DBLP:conf/icde/StonebrakerC05}
\bibfield{author}{\bibinfo{person}{Michael Stonebraker} {and}
  \bibinfo{person}{Ugur {\c{C}}etintemel}.} \bibinfo{year}{2005}\natexlab{}.
\newblock \showarticletitle{{``One size fits all'': an idea whose time has come
  and gone}}. In \bibinfo{booktitle}{{\em 21st International Conference on Data
  Engineering (ICDE'05)}}. \bibinfo{publisher}{IEEE Computer Society},
  \bibinfo{address}{Washington, DC, USA}, \bibinfo{pages}{2--11}.
\newblock
\showISSN{1063-6382}


\bibitem[\protect\citeauthoryear{Storm}{Storm}{2017}]%
        {website:Storm}
Storm.
\newblock \bibinfo{title}{{Apache Storm}}.
\newblock   (\bibinfo{date}{Nov.} \bibinfo{year}{2017}).
\newblock
\showURL{%
Retrieved November 20, 2017 from \url{http://storm.apache.org/}}


\bibitem[\protect\citeauthoryear{Tez}{Tez}{2017}]%
        {website:Tez}
Tez.
\newblock \bibinfo{title}{{Apache Tez}}.
\newblock   (\bibinfo{date}{Nov.} \bibinfo{year}{2017}).
\newblock
\showURL{%
Retrieved November 20, 2017 from \url{http://tez.apache.org/}}


\bibitem[\protect\citeauthoryear{Thusoo, Sarma, Jain, Shao, Chakka, Anthony,
  Liu, Wyckoff, and Murthy}{Thusoo et~al\mbox{.}}{2009}]%
        {hive}
\bibfield{author}{\bibinfo{person}{Ashish Thusoo}, \bibinfo{person}{Joydeep~Sen
  Sarma}, \bibinfo{person}{Namit Jain}, \bibinfo{person}{Zheng Shao},
  \bibinfo{person}{Prasad Chakka}, \bibinfo{person}{Suresh Anthony},
  \bibinfo{person}{Hao Liu}, \bibinfo{person}{Pete Wyckoff}, {and}
  \bibinfo{person}{Raghotham Murthy}.} \bibinfo{year}{2009}\natexlab{}.
\newblock \showarticletitle{Hive: a warehousing solution over a map-reduce
  framework}.
\newblock \bibinfo{journal}{{\em VLDB\/}} (\bibinfo{year}{2009}),
  \bibinfo{pages}{1626--1629}.
\newblock


\bibitem[\protect\citeauthoryear{Trummer and Koch}{Trummer and Koch}{2017}]%
        {trummer2017multi}
\bibfield{author}{\bibinfo{person}{Immanuel Trummer} {and}
  \bibinfo{person}{Christoph Koch}.} \bibinfo{year}{2017}\natexlab{}.
\newblock \showarticletitle{{Multi-objective parametric query optimization}}.
\newblock \bibinfo{journal}{{\em The VLDB Journal\/}} \bibinfo{volume}{26},
  \bibinfo{number}{1} (\bibinfo{year}{2017}), \bibinfo{pages}{107--124}.
\newblock


\bibitem[\protect\citeauthoryear{Vajantri, Toor, and Begoli}{Vajantri
  et~al\mbox{.}}{2017}]%
        {calcite2017polystore}
\bibfield{author}{\bibinfo{person}{Ashwin~Kumar Vajantri},
  \bibinfo{person}{Kunwar Deep~Singh Toor}, {and} \bibinfo{person}{Edmon
  Begoli}.} \bibinfo{year}{2017}\natexlab{}.
\newblock \showarticletitle{{An Apache Calcite-based Polystore Variation for
  Federated Querying of Heterogeneous Healthcare Sources}}. In
  \bibinfo{booktitle}{{\em 2nd Workshop on Methods to Manage Heterogeneous Big
  Data and Polystore Databases}}. \bibinfo{publisher}{IEEE Computer Society},
  \bibinfo{address}{Washington, DC, USA}.
\newblock


\bibitem[\protect\citeauthoryear{Yu, Gadepally, and Stonebraker}{Yu
  et~al\mbox{.}}{2017}]%
        {8091081}
\bibfield{author}{\bibinfo{person}{Katherine Yu}, \bibinfo{person}{Vijay
  Gadepally}, {and} \bibinfo{person}{Michael Stonebraker}.}
  \bibinfo{year}{2017}\natexlab{}.
\newblock \showarticletitle{{Database engine integration and performance
  analysis of the BigDAWG polystore system}}. In \bibinfo{booktitle}{{\em 2017
  IEEE High Performance Extreme Computing Conference (HPEC)}}.
  \bibinfo{publisher}{IEEE Computer Society}, \bibinfo{address}{Washington, DC,
  USA}, \bibinfo{pages}{1--7}.
\newblock


\bibitem[\protect\citeauthoryear{Zaharia, Chowdhury, Franklin, Shenker, and
  Stoica}{Zaharia et~al\mbox{.}}{2010}]%
        {Zaharia:sparksystem}
\bibfield{author}{\bibinfo{person}{Matei Zaharia}, \bibinfo{person}{Mosharaf
  Chowdhury}, \bibinfo{person}{Michael~J. Franklin}, \bibinfo{person}{Scott
  Shenker}, {and} \bibinfo{person}{Ion Stoica}.}
  \bibinfo{year}{2010}\natexlab{}.
\newblock \showarticletitle{{Spark: Cluster Computing with Working Sets}}. In
  \bibinfo{booktitle}{{\em HotCloud}}.
\newblock


\bibitem[\protect\citeauthoryear{Zhou, Larson, and Chaiken}{Zhou
  et~al\mbox{.}}{2010}]%
        {DBLP:conf/icde/ZhouLC10}
\bibfield{author}{\bibinfo{person}{Jingren Zhou},
  \bibinfo{person}{Per{-}{\AA}ke Larson}, {and} \bibinfo{person}{Ronnie
  Chaiken}.} \bibinfo{year}{2010}\natexlab{}.
\newblock \showarticletitle{{Incorporating partitioning and parallel plans into
  the {SCOPE} optimizer}}. In \bibinfo{booktitle}{{\em 2010 IEEE 26th
  International Conference on Data Engineering (ICDE 2010)}}.
  \bibinfo{publisher}{IEEE Computer Society}, \bibinfo{address}{Washington, DC,
  USA}, \bibinfo{pages}{1060--1071}.
\newblock


\end{thebibliography}

\end{document}